\documentclass[useAMS,usenatbib]{mn2e}

\usepackage{natbib}
\usepackage{graphicx}
\usepackage{multicol}
\usepackage{longtable}
\usepackage{amssymb}
\usepackage[width=1.0\textwidth]{caption}
\usepackage[usenames,dvipsnames]{color}
\usepackage{mathtools}
\RequirePackage[pdftex, plainpages = false, pdfpagelabels]{hyperref}
\RequirePackage{subfigure} %
\definecolor{Red}{rgb}{1.0,0,0}

\newcommand{\etal}    {{\it et al}}

\newcommand{\II}      {~{\sc ii}}

\newcommand{\MB}      {Maxwell-Boltzmann}
\newcommand{\hf}      {Hf~2-2}

\title[Coefficients for hydrogen with $\kappa$ distribution]
{Emission and recombination coefficients for hydrogen with $\kappa$-distributed electron energies}

\author[P.J. Storey \& Taha Sochi]
{P.J. Storey$^{1}$, Taha Sochi$^{1}$\thanks{E-mail: t.sochi@ucl.ac.uk} \\
$^{1}$Department of Physics and Astronomy, University College London, Gower Street, London WC1E
6BT, UK}

\begin{document}

\date{Accepted 2014 October 23. Received 2014 October 9; in original form 2014 August 9}

\maketitle

\label{firstpage}

\begin{abstract}
We provide a data set of emission and recombination coefficients of hydrogen using a
$\kappa$-distribution of electron energies rather than the more traditional Maxwell-Boltzmann (MB)
distribution. The data are mainly relevant to thin and relatively cold plasma found in planetary
nebulae and H\II\ regions. The data set extends the previous data sets provided by Storey and
Hummer which were computed using a MB distribution.  The data set, which is placed in the public
domain, is structured as a function of electron number density, temperature and $\kappa$.
Interactive fortran 77 and C++ data servers are also provided as an accessory to probe the data and
obtain Lagrange-interpolated values for any choice of all three variables between the explicitly
computed values.
\end{abstract}

\begin{keywords}
atomic data - atomic processes - radiation mechanisms: general - radiation mechanisms: non-thermal
- ISM: abundances - planetary nebulae: general.
\end{keywords}

\section{Introduction} \label{Introduction}

There are many studies related to the recombination of hydrogen and hydrogenic systems, the most
comprehensive being those of \citet{HummerS87,StoreyH1988} and \citet{StoreyH95}. However, all the
past work was based on a \MB\ distribution of electron energies. There has been, and still is a
general consensus that this is the appropriate distribution for thin nebular plasmas. However, this
was disputed in the past \citep{Hagihara1944} where considerable deviations from the thermodynamic
equilibrium on which the MB relies were claimed although this claim was later discounted
\citep{BohmA1947}.

There has been a recent revival \citep{NichollsDS2012} of the proposal of a non-thermal electron
energy distribution in planetary nebulae and H\II\ regions in the light of the long standing
problem in nebular physics of the contradiction between the results for elemental abundance and
electron temperature as obtained from the optical recombination lines (ORL) and those obtained from
the collisionally excited lines (CEL). According to the recent proposal, the ORL-CEL discrepancy
problem can be resolved by assuming a non-MB electron distribution, specifically the
$\kappa$-distribution. There have been a few recent attempts to assess the merit of this suggestion
\citep{SochiThesis2012,StoreySTemp2013,StoreySHf222014,ZhangLZ2014} using spectroscopic means to
directly sample the free electron energy distribution. With the exception of \cite{StoreySHf222014}
they are all inconclusive, in the sense that the data do not differentiate between a single
$\kappa$-distribution and other models, such as one with two MB components at different
temperatures. \cite{StoreySHf222014} do, however argue that, with a high degree of certainty, the
Balmer line and continuum spectrum of the extreme planetary nebula \hf\ cannot be modeled with a
single $\kappa$-distributed electron energy distribution, while it can be with a model comprising
two MB distributions. It should be noted however that \cite{ZhangLZ2014} analyze the same spectra
of the same object and conclude that either model can adequately model the spectrum. We return to
this apparent contradiction below.

Both \cite{StoreySHf222014} and \cite{ZhangLZ2014} model the Balmer line and continuum spectrum
with MB and $\kappa$ distributions. Typically the continuum intensity is modeled relative to one of
the high Balmer lines, chosen to be close in wavelength to the Balmer edge and apparently
unblended. The continuum spectrum is relatively easy to model with an arbitrary electron energy
distribution but the modeling of the high Balmer line intensities requires a full treatment of the
collisional-radiative recombination process as a function of $\kappa$ as well as the usual density
and temperature variables. \cite{StoreySHf222014} made such a calculation in Case B of
\cite{BakerM1938} and presented some results for the line which they used to normalize intensities,
H11.  Here we publish the full results from those calculations. We note that \cite{ZhangLZ2014}
also use H11 for normalization but they rely on an approximate treatment taken from
\cite{NichollsDS2012} which applies a $\kappa$-dependent scaling function to the H11 emission
coefficient calculated with a MB distribution. \cite{StoreySHf222014} show that this approximation
is poor for the very low values of $\kappa$ that are required to model the \hf\ spectrum, which may
explain why \cite{ZhangLZ2014} reach a different conclusion to \cite{StoreySHf222014}  in the case
of \hf.

The results are provided in two text files where the emission and recombination coefficients are
given as a function of electron number density $N_e$, electron temperature $T_e$ and $\kappa$. We
also provide interactive data servers, in the form of fortran 77 and C++ codes, for mining the data
and obtaining interpolated values in the three variables between the explicitly computed values. In
section~\ref{Model} we give a brief theoretical background about the atomic computational model
used to generate the data, while in section~\ref{Data} we explain the structure of the data set and
provide general instructions and clarifications about how it should be probed and used.
Section~\ref{Conclusions} contains general conclusions and discussions.

\section{Atomic Model}\label{Model}

The fundamental quantity required for the calculation of the hydrogen recombination line
intensities is the coefficient for recombination to a state $nl$ of H. For a free electron energy
distribution $f(E)$, this is given by (e.g. \cite{StoreySHf222014})

\begin{equation}
\begin{multlined}
\alpha(nl) = \frac{R^{\frac{5}{2}}}{\sqrt{2}c^2 m_e^{\frac{3}{2}}} \frac{\omega(nl)}{\omega^+} \\
\int_0^\infty \left( \frac{h\nu}{R} \right)^2 \left( \frac{R}{E} \right)^{\frac{1}{2}}
\sigma(\nu,nl)\ f(E)\ {\rm d}\left(\frac{E}{R} \right) \label{alphaEq}
\end{multlined}
\end{equation}
where $R$ is the Rydberg energy constant, $\omega^+$ and $\omega(nl)$ are the statistical weights
of the initial and final states respectively, $\nu$ is the frequency of the emitted photon, $E$ is
the energy of the free electron, $\sigma(\nu,nl)$ is the cross-section for photoionization which is
the inverse process to recombination, and the other symbols have their standard meanings. The
calculation of the recombination line emission coefficients in a full collisional-radiative
treatment has been described by \citet{HummerS87} and \citet{StoreyH95} and we use the same methods
here.

Traditionally, $f(E)$ is the \MB\ distribution function. Here we use instead the $\kappa$
distribution which is given by \citep{Vasyliunas1968, SummersT91}

\begin{equation}
\begin{multlined}
f_{\kappa}\left(E, T_{\kappa}\right)= \\
\frac{2\sqrt{E}}{\sqrt{\pi(kT_{\kappa})^3}}\frac{\Gamma\left(\kappa+1\right)}{(\kappa-\frac{3}{2})^{\frac{3}{2}}\Gamma\left(\kappa-\frac{1}{2}\right)}\left(1+\frac{E}{(\kappa-\frac{3}{2})
kT_{\kappa}}\right)^{-\left(\kappa+1\right)} \label{kappaEq}
\end{multlined}
\end{equation}
where $\kappa$ is a parameter defining the distribution, $\Gamma$ is the gamma function of the
given arguments, and $T_{\kappa}$ is a temperature characteristic to the particular distribution.
For sufficiently large $\kappa$, $f_{\kappa}(E,T_{\kappa})$ tends to the \MB\ distribution.

For energetically low-lying states of H the dominant processes are recombination and radiative
decay. For higher states, collisional processes become important, with $l$-changing collisions
being the most frequent. In our calculations the $nl$ states of hydrogen are assumed degenerate
with respect to $l$ and in this case the dominant processes that change $l$ are collisions not with
electrons but with H$^+$, He$^+$ and He$^{++}$ ions. In the results described here we retain a \MB\
distribution for these heavier particles. We also retain a \MB\ distribution for the processes that
change energy and $n$, which are dominated by collisions with electrons. Consequently the
emissivities that we calculate should be treated with caution for the high-$n$ states for which
$l$- and $n$-changing collisional processes become important. The boundary of this region is
primarily a function of the electron density, being at approximately $n=100, 75, 50, 30$ and $20$
for densities of $10^2, 10^3, 10^4, 10^5$ and $10^6$~cm$^{-3}$ respectively.

The rate coefficients for $l$-changing collisions used to obtain the above boundary $n$ values were
calculated using the theory described by \cite{PengellyS1964}. Vrinceanu and co-workers have
published a series of papers \citep{Vrinceanu2005, VrinceanuOS2012, VrinceanuOS2014} on electron
and proton induced collisions with Rydberg states of hydrogen. In \citet{VrinceanuOS2012} they
state that the rate coefficients for proton induced $\Delta l=1$ transitions are overestimated by
the theory of \citet{PengellyS1964} by about an order of magnitude. The calculation of the rate
coefficient depends upon an integration of the probability for an $l$-changing transition over the
impact parameter of the incident particle, assumed to travel on a straight-line trajectory. It is
well known that this integral is divergent for $\Delta l = 1$ transitions in a quantum mechanical
treatment. The approximate treatment of the transition probability by \citet{PengellyS1964}
converges on the quantum mechanical result at large impact parameter as illustrated in Figure~1 of
\citet{VrinceanuOS2012}. \citet{PengellyS1964} introduce a cut-off at large impact parameter to
remove the divergence based on collective effects in the plasma or the finite lifetime of the
Rydberg state. The semi-classical approach of \citet{VrinceanuOS2012} does not correctly replicate
the quantum behavior at large impact parameter with the probability instead falling discontinuously
to zero at a finite and relatively small value of the impact parameter. The missing contribution
from large impact parameter is the origin of the order of magnitude difference they report between
their results and those of \citet{PengellyS1964}. We see no reason to prefer their semi-classical
result over the quantum treatment at large impact parameters and therefore consider the
\citet{PengellyS1964} results to be more reliable.

\section{Data} \label{Data}

The $\kappa$-dependent emission coefficients, $\epsilon(N_e,T_e,\kappa)$ are provided in a single
text file called `e1bk.d'. The energy emitted per unit volume per unit time is then $N_e N_+
\epsilon(N_e,T_e,\kappa)$ where $N_+$ is the H$^+$ number density and where all quantities are in
cgs units. The structure of this file is explained in the following bullet points:

\begin{itemize}

\item
\noindent The first row of this file contains (in the following order) the number of $N_e$
values, the number of $T_e$ values and the number of $\kappa$ values for which data are provided.

\item
\noindent The 9 values of $N_e$ are specified by log$_{10}N_e$ = 2.0(0.5)6.0.

\item
\noindent The 16 values of $T_e$ are specified by log$_{10}T_e$ = 2.0(0.2)3.8, 3.9(0.1)4.4.

\item
\noindent The 44 values of $\kappa$ are specified by log$_{10}\kappa$ = 0.20(0.01)0.30,
0.35(0.05)1.0, 1.1(0.1)2.0, 2.2(0.2)3.0, 3.5, 4.0, 5.0, 6.0.

\item
\noindent The data therefore consist of 6336 blocks ($=9\times16\times44$).

\item
\noindent The first row of each block contains information about the block which consists of the
following:
\\
$Z$\hspace{0.5cm}log$_{10}N_e$\hspace{0.5cm}log$_{10}T_e$\hspace{0.5cm}log$_{10}\kappa$\hspace{0.5cm}B\hspace{0.5cm}$n_{min}$\hspace{0.5cm}$n_{max}$ \\
where $Z=1$ is the atomic number of hydrogen, `B' refers to Case B and $n_{min}$ and $n_{max}$ are
the minimum and maximum upper state principal quantum numbers for which emission coefficients are
tabulated. Each block therefore contains 4850 ($=\frac{1}{2}n_m(n_m-1)-1$) entries. These 4850
entries are arranged in 607 rows and hence the total number of rows in each block is 608.

\item
\noindent Of the three variables, $N_e$ $T_e$ and $\kappa$, the fastest varying is $N_e$ followed
by $T_e$ followed by $\kappa$, and hence the ordinal number of a block is given by:

\begin{equation}\label{blockNo}
O_{B}=O_{N}+(O_{T}-1)9+(O_{\kappa}-1)144
\end{equation}
where $O_{B}$, $O_{N}$, $O_{T}$ and $O_{\kappa}$ are the ordinal numbers of block, $N_e$ value,
$T_e$ value and $\kappa$ value respectively. For example the ordinal number of the block for
log$_{10}N_e=4$ ($O_N=5$), log$_{10}T_e=2.6$ ($O_T=4$) and log$_{10}\kappa=0.27$ ($O_{\kappa}=8$)
is: \\
$O_{B}=5+(4-1)9+(8-1)144=1040$ \\
and hence it starts on row 631714 ($=608(O_{B}-1)+2$) and ends on row 632321 ($=608O_{B}+1$).

\item
\noindent The 4850 values of emission coefficients in each block are arranged for transitions from
upper levels $n_u$ to lower levels $n_l$ with $n_u$ in descending order from $n_m$, and $n_l$ in
ascending order from 1 to $(n_u-1)$, and hence the ordinal number for a transition $tr(n_u,n_l)$ is
given by

\begin{equation}\label{traO}
O_{tr}=n_l+\frac{1}{2}(n_m-n_u)(n_m+n_u-1)
\end{equation}
Emission coefficients to $n=1$ are not calculated for Case B and hence are set to zero.

\end{itemize}

A second file named `t1bk.d' contains the hydrogen total recombination coefficients in Case B,
$\alpha(N_e,T_e,\kappa)$, and the total recombination coefficients to the 2s state of hydrogen,
$\alpha_{2s}(N_e,T_e,\kappa)$, such that the number of recombinations per unit volume per unit time
is $N_e N_+ \alpha$ in cgs units. This file contains 12672 entries arranged in 1584 rows. The first
half (6336) of these entries are the total recombination coefficients of hydrogen while the second
half are the total recombination coefficients to the 2s state. The entries in each one of these two
blocks correspond to the 6336 values of physical conditions (i.e. various combinations of $N_e$,
$T_e$ and $\kappa$) positioned according to equation~\ref{blockNo}.

\section{Conclusions and Discussions} \label{Conclusions}

In this paper we computed atomic emission and recombination coefficient data for hydrogen with
electron energies described by a $\kappa$-distribution. The atomic model used in the computation of
these data uses the techniques described by \citet{HummerS87} and \citet{StoreyH95}. The data,
which are placed in the public domain, span ranges of electron number density, temperature and
$\kappa$ useful for modeling and analyzing plasmas such as those found in planetary nebulae and
H\II\ regions. Interactive data servers provide easy access to the data with Lagrange-interpolated
values in all three variables.

\section{Acknowledgment and Statement}

The work of PJS was supported in part by STFC (grant ST/J000892/1). We would like to thank Prof.
Gary Ferland for his constructive review and useful comments. The data described in the present
paper, with the interactive data servers, can be obtained in computer readable format from the
Centre de Donn\'{e}es astronomiques de Strasbourg database. The fortran 77 data server is compiled
and tested thoroughly using gfortran, f77, and Intel fortran compilers on Ubuntu 12.04 and
Scientific Linux platforms, while the C++ server is compiled and tested thoroughly using g++
compiler on Ubuntu 12.04 and Dev-C++ 5.7.1 and Microsoft Visual Studio 6.0 compilers on Windows XP
and Windows 8 64-bit versions. Representative sample results from all these compilers and on all
those platforms are compared and found to be identical within the reported numerical accuracy.

\end{document}